\documentstyle[prd,aps,preprint,tighten,epsfig]{revtex}

\begin{document}

\draft

\title{Model-independent Constraints on the Weak Phase $\alpha$
(or $\phi_2$) and QCD Penguin Pollution in $B \rightarrow \pi\pi$
Decays}
\author{{\bf Zhi-zhong Xing} ~ and ~ {\bf He Zhang}}
\address{CCAST (World Laboratory), P.O. Box 8730,
Beijing 100080, China \\
and Institute of High Energy Physics,
Chinese Academy of Sciences, \\
P.O. Box 918 (4), Beijing 100049, China
\footnote{Mailing address} \\
({\it Electronic address: xingzz@mail.ihep.ac.cn;
zhanghe@mail.ihep.ac.cn}) }
\maketitle

\begin{abstract}
We present an {\it algebraic} isospin approach towards a more
straightforward and model-independent determination of the weak
phase $\alpha$ (or $\phi_2$) and QCD penguin pollution in
$B\rightarrow \pi\pi$ decays. The world averages of current
experimental data allow us to impose some useful constraints on the
isospin parameters of $B\rightarrow \pi\pi$ transitions. We
find that the magnitude of $\alpha$ (or $\phi_2$) extracted
from the indirect CP violation in $\pi^+\pi^-$ mode is in
agreement with the standard-model expectation from other
indirect measurements, but its four-fold discrete ambiguity
has to be resolved in the near future.
\end{abstract}


\newpage

\framebox{\large\bf 1} ~
The major goal of KEK and SLAC $B$-meson factories is to test the
Kobayashi-Maskawa mechanism of CP violation \cite{KM} within the
standard model and to detect possible new sources of CP violation
beyond the standard model. An elegant description of CP violation
in $B$ physics is the unitarity triangle defined by the following
orthogonality relation of six quark mixing matrix elements in the
complex plane \cite{PDG}:
\begin{equation}
V_{ud}V^*_{ub} + V_{cd}V^*_{cb} + V_{td}V^*_{tb} \; =\; 0 \; .
\end{equation}
Three inner angles of this triangle are denoted as
$\alpha$, $\beta$ and $\gamma$ by the BaBar Collaboration, or
equivalently $\phi_1$, $\phi_2$ and $\phi_3$ by the Belle Collaboration:
\begin{eqnarray}
\alpha & \equiv & \phi_2  \equiv
\arg \left ( - \frac{V_{td}V^*_{tb}}{V_{ud}V^*_{ub}} \right ) \; ,
\nonumber \\
\beta & \equiv & \phi_1  \equiv
\arg \left ( - \frac{V_{cd}V^*_{cb}}{V_{td}V^*_{tb}} \right ) \; ,
\nonumber \\
\gamma & \equiv & \phi_3  \equiv
\arg \left ( - \frac{V_{ud}V^*_{ub}}{V_{cd}V^*_{cb}} \right ) \; .
\end{eqnarray}
So far $\beta$ has been rather precisely determined from the
measurement of CP violation in $B^0_d$ vs $\overline{B}^0_d
\rightarrow J/\psi K_{\rm S}$ transitions \cite{2B}, and its value
$\beta \approx 23^\circ$ is compatible very well with the
standard-model expectation. The next experimental step is to
measure $\alpha$ and $\gamma$ to a good degree of accuracy at
$B$-meson factories, such that one may cross-check the
Kobayashi-Maskawa picture of CP violation and probe possible new
physics in the $B$-meson system.

It is well known that $\overline{B}^0_d \rightarrow \pi^+ \pi^-$,
$\overline{B}^0_d \rightarrow \pi^0 \pi^0$ and $B^-_u \rightarrow
\pi^0 \pi^-$ decays can be used to extract the weak angle $\alpha$
in a model-independent way, because the isospin relation of their
transition amplitudes allows us to remove the QCD penguin
pollution \cite{Gronau}. Note, however, that the
experimentally-reported branching fractions of $B\rightarrow
\pi\pi$ decays are all charge-averaged:
\begin{eqnarray}
{\cal B}_{+-} & \equiv & \frac{1}{2} \left [ {\cal B}(B^0_d
\rightarrow \pi^+\pi^-) + {\cal B}(\overline{B}^0_d \rightarrow
\pi^+\pi^-) \right ] \; ,
\nonumber \\
{\cal B}_{00} ~ & \equiv & \frac{1}{2} \left [ {\cal B}(B^0_d
\rightarrow \pi^0\pi^0) + {\cal B}(\overline{B}^0_d \rightarrow
\pi^0\pi^0) \right ] \; ,
\nonumber \\
{\cal B}_{0\pm} ~ & \equiv & \frac{1}{2} \left [
{\cal B}(B^+_u \rightarrow \pi^0\pi^+)
+ {\cal B}(B^-_u \rightarrow \pi^0\pi^-) \right ] \; .
\end{eqnarray}
The world averages of current BaBar \cite{BaBar}, Belle
\cite{Belle} and CLEO \cite{CLEO} data on ${\cal B}_{+-}$, ${\cal
B}_{00}$ and ${\cal B}_{0\pm}$ are listed in Table 1 \cite{WA}. In
addition, the {\it direct} CP-violating asymmetries between
$\overline{B}^0_d \rightarrow \pi^+ \pi^-$, $\overline{B}^0_d
\rightarrow \pi^0 \pi^0$, $B^-_u \rightarrow \pi^0 \pi^-$ and
their CP-conjugate decays can be defined as
\begin{eqnarray}
{\cal C}_{+-} & \equiv & \frac{{\cal B}(B^0_d \rightarrow
\pi^+\pi^-) - {\cal B}(\overline{B}^0_d \rightarrow \pi^+\pi^-)}
{{\cal B}(B^0_d \rightarrow \pi^+\pi^-) + {\cal
B}(\overline{B}^0_d \rightarrow \pi^+\pi^-)} \;\; ,
\nonumber \\
{\cal C}_{00} ~ & \equiv & \frac{{\cal B}(B^0_d \rightarrow
\pi^0\pi^0) - {\cal B}(\overline{B}^0_d \rightarrow \pi^0\pi^0)}
{{\cal B}(B^0_d \rightarrow \pi^0\pi^0) + {\cal
B}(\overline{B}^0_d \rightarrow \pi^0\pi^0)} \;\; ,
\nonumber \\
{\cal A}_{0\pm} & \equiv & \frac{{\cal B}(B^-_u \rightarrow \pi^0\pi^-)
- {\cal B}(B^+_u \rightarrow \pi^0\pi^+)}
{{\cal B}(B^-_u \rightarrow \pi^0\pi^-)
+ {\cal B}(B^+_u \rightarrow \pi^0\pi^+)} \;\; .
\end{eqnarray}
The world averages of current BaBar \cite{BaBar} and Belle
\cite{Belle} data on ${\cal C}_{+-}$, ${\cal C}_{00}$ and ${\cal
A}_{0\pm}$ are also shown in Table 1 \cite{WA}. These two
collaborations have actually measured the time-dependent rates of
$B^0_d$ vs $\overline{B}^0_d \rightarrow \pi^+\pi^-$ decays on the
$\Upsilon (4S)$ resonance:
\begin{eqnarray}
\Gamma [B^0_d(\Delta t) \rightarrow \pi^+\pi^-] & = &
\frac{e^{-|\Delta t|/\tau^{~}_0}}{4\tau^{~}_0} \left [ 1 -
{\cal S}_{+-} \sin \left (\Delta m^{~}_d \Delta t \right ) + {\cal C}_{+-}
\cos \left (\Delta m^{~}_d \Delta t \right ) \right ] \; ,
\nonumber \\
\Gamma [\overline{B}^0_d(\Delta t) \rightarrow \pi^+\pi^-] & = &
\frac{e^{-|\Delta t|/\tau^{~}_0}}{4\tau^{~}_0} \left [ 1 + {\cal
S}_{+-} \sin \left (\Delta m^{~}_d \Delta t \right ) - {\cal
C}_{+-} \cos \left (\Delta m^{~}_d \Delta t \right ) \right ] \; ,
\end{eqnarray}
where $\tau^{~}_0$ is the lifetime of neutral $B$ mesons, and
${\cal S}_{+-}$ signifies the {\it indirect} CP violation arising
from the interplay between decay and $B^0_d$-$\overline{B}^0_d$
mixing \cite{BS}. A similar time-dependent measurement can be done
for $B^0_d$ vs $\overline{B}^0_d \rightarrow \pi^0\pi^0$ decays,
whose rates consist of ${\cal C}_{00}$ and ${\cal S}_{00}$
corresponding to ${\cal C}_{+-}$ and ${\cal S}_{+-}$ in Eq. (5).
Only ${\cal S}_{+-}$ has been determined from the KEK and SLAC
experiments, and its world average \cite{WA} is given in Table 1.

Although the relevant experimental data on direct CP violation remain
quite preliminary, they can be used to do a quantitative
analysis of $B\rightarrow \pi\pi$ decays. The main purpose of this paper
is to recommend an {\it algebraic} isospin approach, which allows us to
figure out the ranges of $B\rightarrow \pi\pi$ isospin parameters and
to determine the weak phase $\alpha$ (or $\phi_2$) from ${\cal S}_{+-}$
and (or) ${\cal S}_{00}$ 
in a more straightforward and model-independent way.
We find that the allowed region of $\alpha$ is in agreement with the
standard-model expectation from other indirect measurements, but its
four-fold discrete ambiguity has to be resolved in the near future.

\vspace{0.3cm}

\framebox{\large 2} ~
Under isospin symmetry and in the neglect of electroweak
penguin contributions \cite{Rosner}, the amplitudes of
$B^0_d \rightarrow \pi^+ \pi^-$,
$B^0_d \rightarrow \pi^0 \pi^0$ and
$B^+_u \rightarrow \pi^0 \pi^+$ decays (or their CP-conjugate
processes) form a triangle in the complex plane:
\begin{eqnarray}
A(B^0_d \rightarrow \pi^+\pi^-) + \sqrt{2}
A(B^0_d \rightarrow \pi^0\pi^0) & = &
\sqrt{2} A(B^+_u \rightarrow \pi^0 \pi^+) \; ,
\nonumber \\
A(\overline{B}^0_d \rightarrow \pi^+\pi^-) + \sqrt{2}
A(\overline{B}^0_d \rightarrow \pi^0\pi^0) & = & \sqrt{2} A(B^-_u
\rightarrow \pi^0 \pi^-) \; .
\end{eqnarray}
The magnitudes of $A(B^-_u \rightarrow \pi^0 \pi^-)$ and $A(B^+_u
\rightarrow \pi^0 \pi^+)$ are identical to each other in this safe
approximation \cite{Gronau,Iso}, hence the CP-violating asymmetry
${\cal A}_{0\pm}$ vanishes. Table 1 indicates that the
experimental data are in good agreement with the expectation of
${\cal A}_{0\pm} \approx 0$. In terms of the charge-averaged
branching fractions in Eq. (3) and the direct CP-violating
asymmetries in Eq. (4), let us follow Ref. \cite{Xing03} to
explicitly express the parameters
\begin{eqnarray}
r & = & |r| e^{i\theta} \; \equiv \; \frac{A_0}{A_2} \; ,
\nonumber \\
\overline{r} & = & |\overline{r}| e^{i\overline{\theta}} \; \equiv
\; \frac{\overline{A}_0}{\overline{A}_2} \; ,
\end{eqnarray}
which stand for the ratios of $I=0$ and $I=2$ isospin amplitudes
in $B^0_d \rightarrow \pi^+\pi^-$ (or $\pi^0\pi^0$) and
$\overline{B}^0_d \rightarrow \pi^+\pi^-$ (or $\pi^0\pi^0$)
decays. The results are
\begin{eqnarray}
|r| & = & \frac{\sqrt{3 {\cal B}_{+-} \left (1 + {\cal C}_{+-} \right )
+ 3 {\cal B}_{00} \left (1 + {\cal C}_{00} \right ) - 2 \kappa
{\cal B}_{0\pm}}}{\sqrt{\kappa {\cal B}_{0\pm}}} \;\; ,
\nonumber \\
|\overline{r}| & = & \frac{\sqrt{3 {\cal B}_{+-} \left (1 - {\cal
C}_{+-} \right ) + 3 {\cal B}_{00} \left (1 - {\cal C}_{00} \right
) - 2 \kappa {\cal B}_{0\pm}}}{\sqrt{\kappa {\cal B}_{0\pm}}} \;\;
;
\end{eqnarray}
and
\begin{eqnarray}
\theta & = & \pm \arccos \left [
\frac{6 {\cal B}_{00} \left (1 + {\cal C}_{00} \right )
- 3 {\cal B}_{+-} \left (1 + {\cal C}_{+-} \right ) - 2 \kappa
{\cal B}_{0\pm}}{4 \sqrt{\kappa {\cal B}_{0\pm} \left [
3 {\cal B}_{+-} \left (1 + {\cal C}_{+-} \right )
+ 3 {\cal B}_{00} \left (1 + {\cal C}_{00} \right ) - 2 \kappa
{\cal B}_{0\pm} \right ]}} \right ] \; ,
\nonumber \\
\overline{\theta} & = & \pm \arccos \left [ \frac{6 {\cal B}_{00}
\left (1 - {\cal C}_{00} \right ) - 3 {\cal B}_{+-} \left (1 -
{\cal C}_{+-} \right ) - 2 \kappa {\cal B}_{0\pm}}{4 \sqrt{\kappa
{\cal B}_{0\pm} \left [ 3 {\cal B}_{+-} \left (1 - {\cal C}_{+-}
\right ) + 3 {\cal B}_{00} \left (1 - {\cal C}_{00} \right ) - 2
\kappa {\cal B}_{0\pm} \right ]}} \right ] \; ,
\end{eqnarray}
where $\kappa \equiv \tau^{~}_0/\tau_{\pm} = 0.921 \pm 0.017$
\cite{PDG} denotes the lifetime ratio of neutral and charged $B$
mesons. Eqs. (8) and (9) clearly show that $\overline{r} = r$
(i.e., $|\overline{r}| = |r|$ and $\overline{\theta} = \theta$)
would hold, if ${\cal C}_{+-} = {\cal C}_{00} =0$ held. Hence the
deviation of $\overline{r}$ from $r$ is a measure of direct CP
violation in $B^0_d$ vs $\overline{B}^0_d \rightarrow \pi^+\pi^-$
and $\pi^0\pi^0$ decays
\footnote{It is worth mentioning that the difference between $\phi
\equiv \arg [A(\overline{B}^0_d \rightarrow \pi^+\pi^-)/
A(\overline{B}^0_d \rightarrow \pi^0\pi^0)]$ and $\varphi \equiv
\arg [A(B^0_d \rightarrow \pi^+\pi^-)/ A(B^0_d \rightarrow
\pi^0\pi^0)]$ measures the existence of direct CP violation too
\cite{Xing03}. With the help of Eqs. (3), (4) and (6), one may
obtain
\begin{eqnarray}
\cos\phi & = & \frac{2 \kappa {\cal B}_{0\pm} - {\cal B}_{+-}
\left (1 - {\cal C}_{+-} \right ) -
2 {\cal B}_{00} \left (1 - {\cal C}_{00} \right )}
{2 \sqrt{2 {\cal B}_{+-} {\cal B}_{00} \left (1 - {\cal C}_{+-} \right )
\left (1 - {\cal C}_{00} \right )}} \; ,
\nonumber \\
\cos\varphi & = & \frac{2 \kappa {\cal B}_{0\pm} - {\cal B}_{+-}
\left (1 + {\cal C}_{+-} \right ) -
2 {\cal B}_{00} \left (1 + {\cal C}_{00} \right )}
{2 \sqrt{2 {\cal B}_{+-} {\cal B}_{00} \left (1 + {\cal C}_{+-} \right )
\left (1 + {\cal C}_{00} \right )}} \; .
\nonumber
\end{eqnarray}
Obviously, $\cos\varphi = \cos\phi$ would hold if both ${\cal C}_{+-}$
and ${\cal C}_{00}$ were vanishing.
Note that the notations of direct CP-violating
asymmetries in Ref. \cite{Xing03} are ${\cal A}_{+-} = -{\cal C}_{+-}$
and ${\cal A}_{00} = -{\cal C}_{00}$.}.

The CP-violating parameters ${\cal S}_{+-}$ and ${\cal S}_{00}$ in
Eq. (5) are related to the weak angle $\alpha$ as follows
\cite{Xing03}:
\begin{eqnarray}
{\cal S}_{+-} & = & (1 + {\cal C}_{+-}) {\rm Im} \left [
\frac{q}{p} \cdot \frac{A(\overline{B}^0_d \rightarrow
\pi^+\pi^-)} {A(B^0_d \rightarrow \pi^+\pi^-)} \right ] = (1 +
{\cal C}_{+-}) |R| \sin [ 2 (\alpha + \Theta)] \; ,
\nonumber \\
{\cal S}_{00} ~ & = & (1 + {\cal C}_{00}) {\rm Im} \left [
\frac{q}{p} \cdot \frac{A(\overline{B}^0_d \rightarrow
\pi^0\pi^0)} {A(B^0_d \rightarrow \pi^0\pi^0)} \right ] = (1 +
{\cal C}_{00}) |\overline{R}| \sin [ 2 (\alpha +
\overline{\Theta})] \; ,
\end{eqnarray}
where $q/p \approx (V_{td}V^*_{tb})/(V^*_{td}V_{tb})$ denotes the
weak phase of $B^0_d$-$\overline{B}^0_d$ mixing \cite{BS}, and
\begin{eqnarray}
R & = & |R| e^{2i\Theta} \; =\; \frac{1 - \overline{r}}{1-r} \; ,
\nonumber \\
\overline{R} & = & |\overline{R}| e^{2i\overline{\Theta}} \; =\;
\frac{2+ \overline{r}}{2+r} \; .
\end{eqnarray}
To be specific, we have
\begin{eqnarray}
|R| & = & \frac{\sqrt{1 - 2|\overline{r}|\cos\overline{\theta} +
|\overline{r}|^2}} {\sqrt{1 - 2|r|\cos\theta + |r|^2}} \; ,
\nonumber \\
|\overline{R}| & = & \frac{\sqrt{4 +
4|\overline{r}|\cos\overline{\theta} + |\overline{r}|^2}} {\sqrt{4
+ 4|r|\cos\theta + |r|^2}} \; ;
\end{eqnarray}
and
\begin{eqnarray}
\Theta & = & +\frac{1}{2}\arctan \left [ \frac{|r|\sin\theta -
|\overline{r}|\sin\overline{\theta} - |r| |\overline{r}| \sin
(\theta - \overline{\theta})} {1 - |r|\cos\theta -
|\overline{r}|\cos\overline{\theta} + |r| |\overline{r}| \cos
(\theta - \overline{\theta})} \right ] \; ,
\nonumber \\
\overline{\Theta} & = & -\frac{1}{2}\arctan \left [
\frac{2|r|\sin\theta - 2|\overline{r}|\sin\overline{\theta} + |r|
|\overline{r}| \sin (\theta - \overline{\theta})} {4 +
2|r|\cos\theta + 2|\overline{r}|\cos\overline{\theta} + |r|
|\overline{r}| \cos (\theta - \overline{\theta})} \right ] \; .
\end{eqnarray}
If there were no direct CP violation in $B\rightarrow \pi\pi$
transitions (i.e., ${\cal C}_{+-} = {\cal C}_{00} =0$ or
$\overline{r} =r$), we would arrive at $\overline{R} = R =1$ from
Eqs. (11)--(13). In this case, Eq. (10) would be simplified to
${\cal S}_{+-} = {\cal S}_{00} = \sin 2\alpha$. It is therefore
necessary to pin down ${\cal C}_{+-}$ and ${\cal C}_{00}$ to a
reasonable degree of accuracy, in order to extract the weak angle
$\alpha$ from ${\cal S}_{+-}$ and (or) ${\cal S}_{00}$.

\vspace{0.3cm}

\framebox{\large 3} ~ Now we carry out a numerical analysis of
$B\rightarrow \pi\pi$ decays and CP violation by using the isospin
formulas obtained above and current experimental data listed in
Table 1. The 1$\sigma$, 2$\sigma$ and 3$\sigma$ confidence regions
of $(|r|, |\overline{r}|)$, $(\cos\theta, \cos\overline{\theta})$,
$(|R|, |\overline{R}|)$, $(\Theta, \overline{\Theta})$ and
$(\alpha, {\cal S}_{+-})$ are shown in Figs. 1--3, while the
central (best-fit) values of these parameters are given in Table
2. Some discussions are in order.

(1) The moduli $|r|$, $|\overline{r}|$, $|R|$ and $|\overline{R}|$
can be determined without any discrete ambiguity (see Fig. 1 for
illustration). The difference between $|r|$ and $|\overline{r}|$
is quite obvious, although it remains possible for $|\overline{r}|
= |r|$ to hold at the 3$\sigma$ level. In contrast, it is likely
to have $|\overline{R}| = |R|$ at the 1$\sigma$ level, but their
best-fit values are different from each other. Direct CP violation
is therefore expected to manifest itself in $B\rightarrow \pi\pi$
transitions.

(2) Although $\cos\theta$ and $\cos\overline{\theta}$ can be
uniquely determined, $\theta$ or $\overline{\theta}$ involves
two-fold ambiguity. The difference between $\theta$ and
$\overline{\theta}$ is apparent, but $\cos\overline{\theta} =
\cos\theta$ remains possible at the 1$\sigma$ level. More precise
experimental data will allow us to fix $r$ and $\overline{r}$,
both their moduli and their phases, to a better degree of
accuracy.

(3) The two-fold ambiguity of $\theta$ or $\overline{\theta}$
leads to the four-fold ambiguity of $\Theta$ or
$\overline{\Theta}$, as illustrated in Fig. 2 and Table 2. An
interesting feature of our results is that the signs of $\Theta$
and $\overline{\Theta}$ are essentially opposite. Although it is
possible to have $\overline{\Theta} = \Theta$ at the 2$\sigma$
level for case (A) or (D), we find that $\overline{\Theta} \neq
\Theta$ holds even at the 3$\sigma$ level for case (B) or (C). In
particular, the possibility of $\Theta =0$ and (or)
$\overline{\Theta} =0$ is strongly disfavored, implying the
presence of QCD penguin pollution or direct CP violation in
$B\rightarrow \pi\pi$ decay modes. The typical values of $\Theta$
and $\overline{\Theta}$ are $\Theta \sim \pm 19^\circ$ and
$\overline{\Theta} \sim \mp 26^\circ$, as shown in Table 2.

(4) Because of the four-fold ambiguity associated with $\Theta$ or
$\overline{\Theta}$, the result of $\alpha$ determined from ${\cal
S}_{+-}$ involves the four-fold discrete ambiguity too, as
illustrated in Fig. 3. Table 2 tells us that the central values of
$\alpha$ can be $\alpha \sim 122^\circ$ (A), $135^\circ$ (B),
$86^\circ$ (C) or $95^\circ$ (D). This result is certainly in
agreement with the standard-model expectation of $\alpha$ (i.e.,
$\alpha \sim 90^\circ$) from other indirect measurements
\cite{PDG}. Once the CP-violating asymmetry ${\cal S}_{00}$ is
also measured, it will be possible to completely or partly remove
the discrete ambiguity of $\alpha$ \cite{Gronau}. Then one may
constrain the weak phase $\alpha$ and QCD penguin pollution in
$B\rightarrow \pi\pi$ decays at a much better confidence level.

It is worth remarking that the validity of our isospin analysis
relies on the assumption of negligible electroweak penguin effects.
The electroweak penguin contribution to $B\rightarrow \pi\pi$
decays is in general expected to be insignificant \cite{Rosner}.
This expectation would be problematic or incorrect, if
${\cal A}_{0\pm} \neq 0$ were experimentally established \cite{Xing03}.
Note also that final-state interactions in $B\rightarrow \pi\pi$
transitions consist of both elastic
$\pi\pi \rightleftharpoons \pi\pi$ rescattering and some
possible inelastic rescattering effects
\footnote{The transition amplitudes of $B^0_d \rightarrow \pi^+
\pi^-$, $B^0_d \rightarrow \pi^0 \pi^0$ and $B^+_u \rightarrow
\pi^0 \pi^+$ decay modes (or their CP-conjugate processes) may
still form an isospin triangle in the complex plane, even if the
inelastic $\pi\pi \rightleftharpoons D\overline{D}$ rescattering
effects are taken into account \cite{Xing00}. In this complicated
case, however, a model-independent determination of $\alpha$ from
${\cal S}_{+-}$ and ${\cal S}_{00}$ would be rather difficult
\cite{X95}.}.
Whether the latter is negligibly small or not remains an open
question \cite{Donoghue}. To answer this question requires more precise
measurements of both branching fractions and CP-violating asymmetries of
$B\rightarrow \pi\pi$ decays.

\vspace{0.3cm}

\framebox{\large 4} ~
We have presented an {\it algebraic} isospin analysis of
rare $B\rightarrow \pi\pi$ decays by taking account of the fact
that the experimentally-reported branching fractions are
charge-averaged and large direct CP violation may exist in them.
This approach is more straightforward than the originally-proposed
{\it geometric} approach, from which the weak phase $\alpha$ (or
$\phi_2$) and QCD penguin pollution are determined through the
reconstruction of two isospin triangles. Therefore, our method is
expected to be very useful to analyze the future experimental data
on $B\rightarrow \pi\pi$ transitions and CP violation in a
model-independent way.

Although the present experimental data (in particular, those on direct
and indirect CP violation in $B\rightarrow \pi\pi$ decays) are not
sufficiently precise, they can impose some instructive constraints
on the parameter space of QCD penguin effects. Furthermore, we find
that the allowed region of $\alpha$ (or $\phi_2$) is actually in
agreement with the standard-model expectation from other indirect
measurements. To resolve the four-fold discrete ambiguity associated
with the magnitude of $\alpha$ (or $\phi_2$) determined from the indirect
CP-violating asymmetry in $\pi^+\pi^-$ mode, a  measurement of the
similar CP-violating asymmetry in $\pi^0\pi^0$ mode is necessary.
We expect that more accurate measurements of such charmless $B$
decays will help us to test the consistency of the Kobayashi-Maskawa
mechanism of CP violation 
\footnote{For example, the relationship
$\sin\alpha/\sin\beta = |V_{cd}/V_{ud}|/|V_{ub}/V_{cb}|$ \cite{WX}, 
which holds as a straightforward result of the unitarity-triangle 
defined in Eq. (1), can be numerically tested with more accurate 
data of $\alpha$ and $|V_{ub}/V_{cb}|$.}
and to probe possible new physics beyond the
standard model.

\vspace{0.4cm}

This work was supported in part by the National
Nature Science Foundation of China.

\newpage

\begin{table}
\caption{The world averages of current experimental data on the
charge-averaged branching fractions (${\cal B}_{+-}$, ${\cal B}_{00}$,
${\cal B}_{0\pm}$), direct CP-violating asymmetries (${\cal C}_{+-}$,
${\cal C}_{00}$, ${\cal A}_{0\pm}$) and indirect CP-violating
asymmetries (${\cal S}_{+-}$, ${\cal S}_{00}$) of $B\rightarrow \pi\pi$
decays [8].}
\begin{center}
\begin{tabular}{l|l}
& World average  \\ \hline
${\cal B}_{+-}$
& $(4.6 \pm 0.4) \times 10^{-6}$
\\
${\cal B}_{00}$
& $(1.51 \pm 0.28) \times 10^{-6}$
\\
${\cal B}_{0\pm}$
& $(5.5 \pm 0.6) \times 10^{-6}$
\\ \hline
${\cal C}_{+-}$
& $-0.37 \pm 0.11$
\\
${\cal C}_{00}$
& $-0.28 \pm 0.39$
\\
${\cal A}_{0\pm}$
& $-0.02 \pm 0.07$
\\ \hline
${\cal S}_{+-}$
& $-0.61 \pm 0.14$
\\
${\cal S}_{00}$
& ---
\end{tabular}
\end{center}
\end{table}

\begin{table}
\caption{The central values of eight isospin parameters ($|r|$,
$|\overline{r}|$; $\theta$, $\overline{\theta}$; $|R|$,
$|\overline{R}|$; $\Theta$, $\overline{\Theta}$) and the weak
phase $\alpha$ constrained from the world averages of current
BaBar and Belle data on $B\rightarrow \pi\pi$ decays [8], where we
have taken into account the two-fold ambiguity associated with
$\theta$ and $\overline{\theta}$ as well as the four-fold
ambiguity associated with $\Theta$, $\overline{\Theta}$ and
$\alpha$.}
\begin{center}
\begin{tabular}{c|llll}
& Case (A) & Case (B) & Case (C) & Case (D)  \\ \hline
$|r|$
& $0.6$
& $0.6$
& $0.6$
& $0.6$
\\
$|\overline{r}|$ & $1.7$ & $1.7$ & $1.7$ & $1.7$
\\
$\theta$
& $+180^\circ$
& $-180^\circ$
& $+180^\circ$
& $-180^\circ$
\\
$\overline{\theta}$ & $+120^\circ$ & $+120^\circ$ & $-120^\circ$ &
$-120^\circ$
\\ \hline
$|R|$
& $1.5$
& $1.5$
& $1.5$
& $1.5$
\\
$|\overline{R}|$ & $1.3$ & $1.3$ & $1.3$ & $1.3$
\\
$\Theta$
& $-18^\circ$
& $-20^\circ$
& $+20^\circ$
& $+18^\circ$
\\
$\overline{\Theta}$ & $+25^\circ$ & $+27^\circ$ & $-27^\circ$ &
$-25^\circ$
\\ \hline
$\alpha$
& $122^\circ$
& $135^\circ$
& $86^\circ$
& $95^\circ$
\end{tabular}
\end{center}
\end{table}

\newpage

\begin{figure}[t]
\vspace{-1.7cm}
\epsfig{file=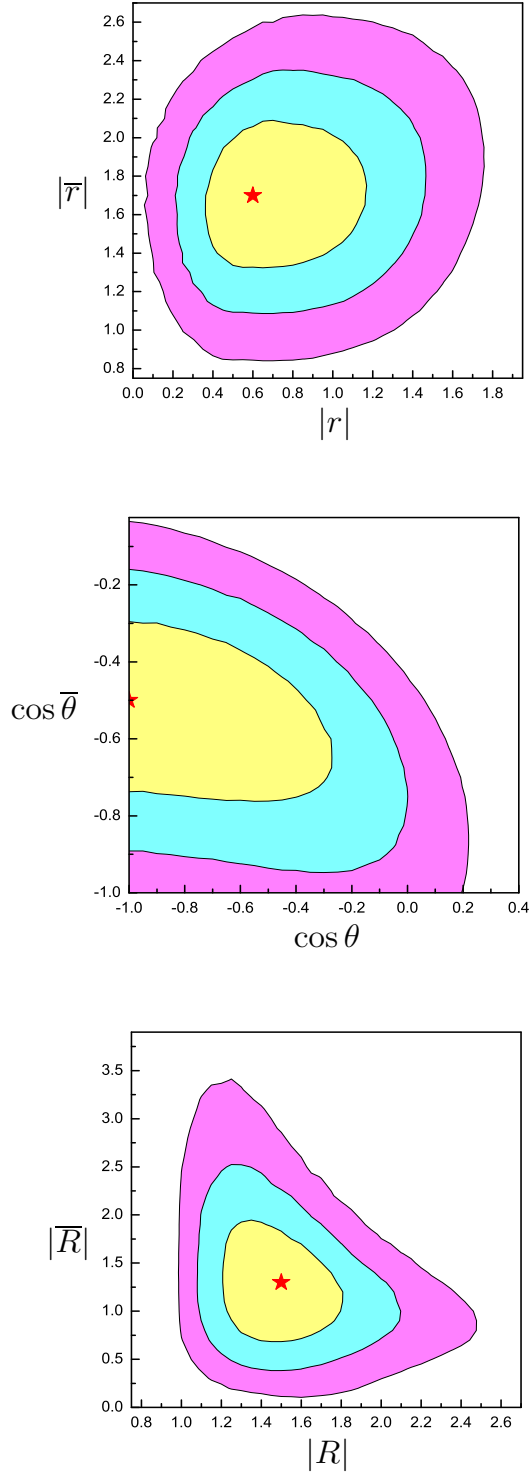,bbllx=2.3cm,bblly=4.3cm,bburx=18cm,bbury=30cm,%
width=14cm,height=22cm,angle=0,clip=0}
\vspace{0.3cm}
\caption{The 1$\sigma$, 2$\sigma$ and 3$\sigma$ confidence regions
of $(|r|, |\overline{r}|)$, $(\cos\theta, \cos\overline{\theta})$
and $(|R|, |\overline{R}|)$ parameters, constrained by the isospin
relations and current experimental data.}
\end{figure}

\newpage

\begin{figure}[t]
\vspace{-1.7cm}
\epsfig{file=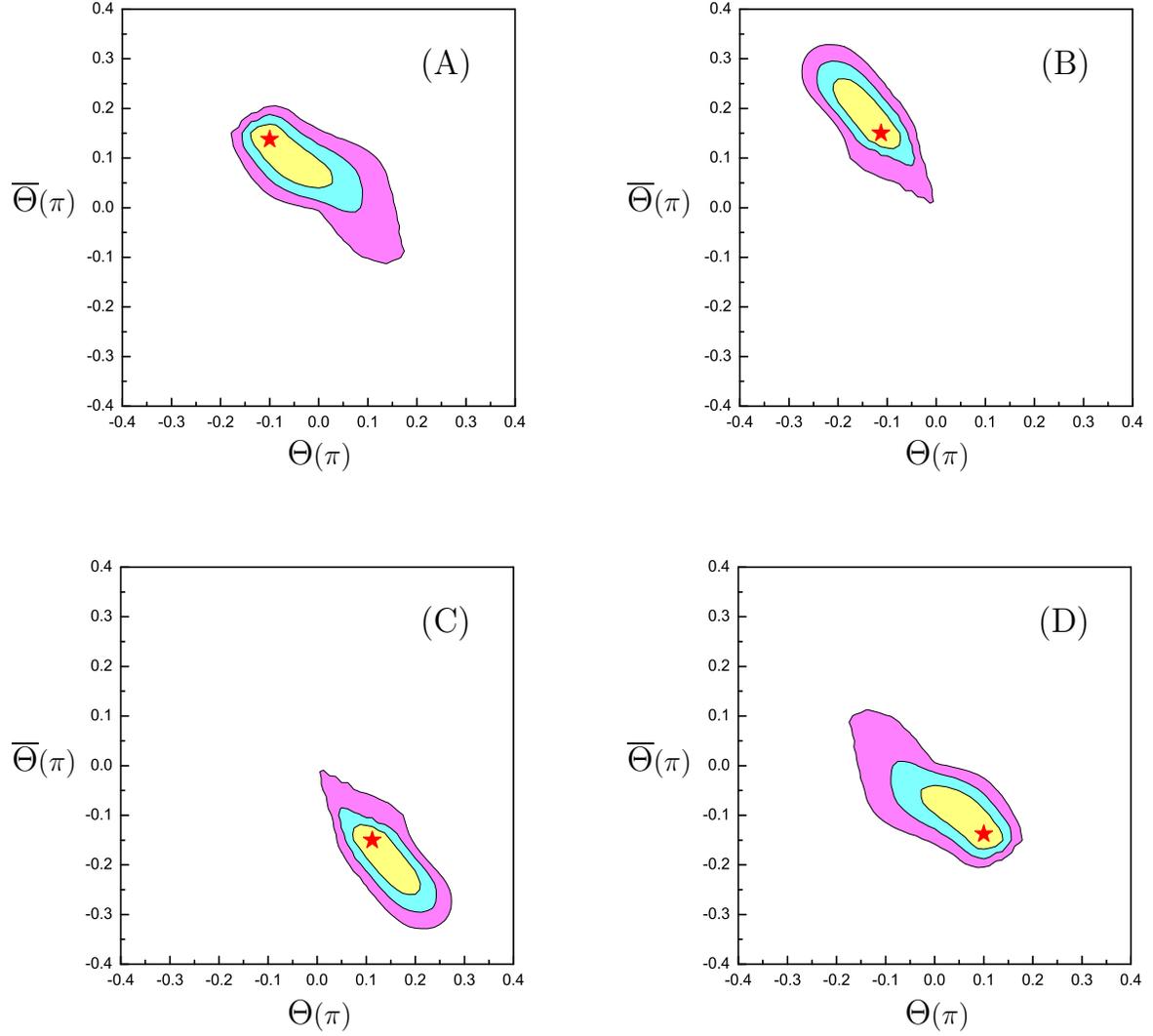,bbllx=1.8cm,bblly=4.3cm,bburx=17cm,bbury=28cm,%
width=14cm,height=22cm,angle=0,clip=0}
\vspace{-4.1cm}
\caption{The 1$\sigma$, 2$\sigma$ and 3$\sigma$ confidence regions
of $\Theta$ and $\overline{\Theta}$ with four-fold discrete ambiguity,
obtained from the isospin analysis of current experimental data.}
\end{figure}

\newpage

\begin{figure}[t]
\vspace{-1.7cm}
\epsfig{file=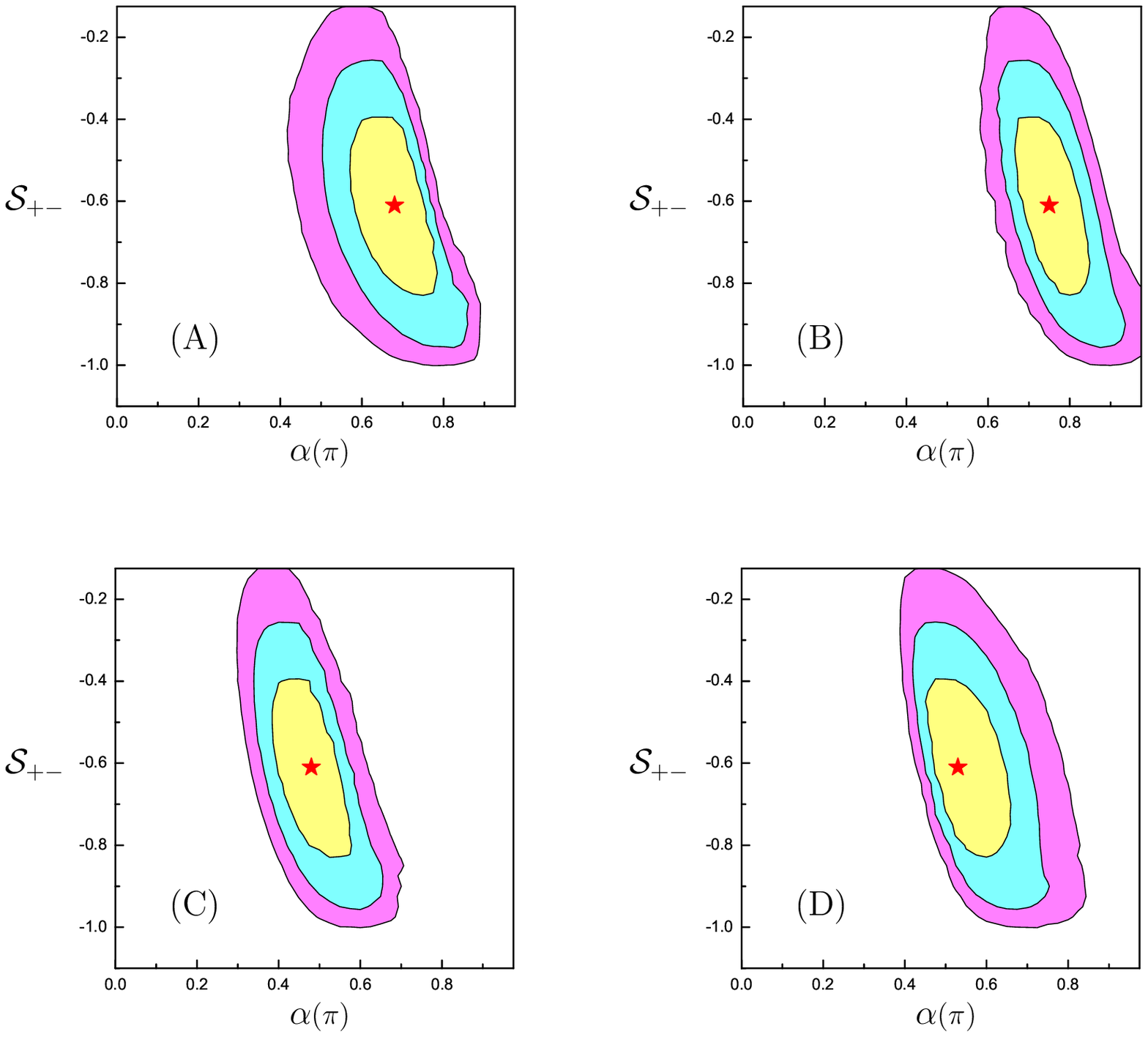,bbllx=1.8cm,bblly=4.3cm,bburx=17cm,bbury=28cm,%
width=14cm,height=22cm,angle=0,clip=0} \vspace{-4.1cm}
\caption{The 1$\sigma$, 2$\sigma$ and 3$\sigma$ confidence regions
of $\alpha$ with four-fold discrete ambiguity, extracted from
current experimental data on ${\cal S}_{+-}$.}
\end{figure}

\newpage

\begin{thebibliography}{99}
\bibitem{KM} M. Kobayashi and T. Maskawa,
Prog. Theor. Phys. {\bf 49}, 652 (1973).

\bibitem{PDG} Particle Data Group, S. Eidelman {\it et al.},
Phys. Lett. B {\bf 592}, 1 (2004).

\bibitem{2B} BaBar Collaboration, B. Aubert {\it et al.},
Phys. Rev. Lett. {\bf 87}, 091801 (2001);
Belle Collaboration, K. Abe {\it et al.},
Phys. Rev. Lett. {\bf 87}, 091802 (2001).

\bibitem{Gronau} M. Gronau and D. London,
Phys. Rev. Lett. {\bf 65}, 3381 (1990).

\bibitem{BaBar} BaBar Collaboration, B. Aubert {\it et al.},
hep-ex/0408081; hep-ex/0408089; BaBar-Conf-04/047; hep-ex/0412037;
and references therein.

\bibitem{Belle} Belle Collaboration, K. Abe {\it et al.},
Phys. Rev. Lett. {\bf 93}, 021601 (2004);
hep-ex/0408101;
Y. Chao {\it et al.}, Phys. Rev. Lett. {\bf 93}, 191802 (2004);
and references therein.

\bibitem{CLEO} CLEO Collaboration, A. Bornheim {\it et al.},
Phys. Rev. D {\bf 68}, 052002 (2003); and references therein.

\bibitem{WA} Heavy Flavor Averaging Group, J. Alexander {\it et al.},
hep-ex/0412073.

\bibitem{BS} I.I. Bigi and A.I. Sanda, {\it CP Violation},
Cambridge University Press (2000).

\bibitem{Rosner} N.G. Deshpande and X.G. He,
Phys. Rev. Lett. {\bf 74}, 26 (1995); M. Gronau, O.F.
Hern$\rm\acute{a}$ndez, D. London, and J.L. Rosner, Phys. Rev. D
{\bf 52}, 6374 (1995);
R. Fleischer, Phys. Rept. {\bf 370}, 537 (2002).

\bibitem{Iso} Y. Grossman and H.R. Quinn,
Phys. Rev. D {\bf 58}, 017504 (1998);
W.S. Hou and K.C. Yang, Phys. Rev. Lett. {\bf 84}, 4806 (2000);
M. Gronau and J.L. Rosner, Phys. Rev. D {\bf 65}, 093012 (2002);
Y.Y. Charng and H.N. Li, Phys. Lett. B {\bf 594}, 185 (2004);
and references therein.

\bibitem{Xing03} Z.Z. Xing, Phys. Rev. D {\bf 68}, 071301 (2003);
D. Du and Z.Z. Xing, Phys. Rev. D {\bf 47}, 2825 (1993).

\bibitem{Xing00} Z.Z. Xing, Phys. Lett. B {\bf 493}, 301 (2000);
and references therein.

\bibitem{X95} C. Hamzaoui and Z.Z. Xing, Phys. Lett. B {\bf 360},
131 (1995).

\bibitem{Donoghue} J.F. Donoghue, E. Golowich, A.A. Petrov, and
J.M. Soares, Phys. Rev. Lett. {\bf 77}, 2178 (1996);
Z.Z. Xing, Phys. Rev. D {\bf 53}, 2847 (1996).

\bibitem{WX} See, e.g., D.D. Wu and Z.Z. Xing,
Phys. Lett. B {\bf 341}, 386 (1995);
H. Fritzsch and Z.Z. Xing, Nucl. Phys. B {\bf 556}, 49 (1999). 
\end{thebibliography}
\end{document}